\documentclass[%
aps,
twocolumn,
superscriptaddress,
nofootinbib,
 amsmath,amssymb,
prd,
nopacs,
longbibliography
]{revtex4-2}

\usepackage[colorlinks,linkcolor=blue,citecolor=blue,urlcolor=blue,hyperindex,driverfallback=dvipdfm]{hyperref}

\usepackage{upgreek}
\usepackage{amsmath,amssymb,amsfonts}
\usepackage{appendix}
\usepackage{bm}
\usepackage{color}
\usepackage[normalem]{ulem} 

\usepackage{textcomp}

\usepackage{graphicx}
\usepackage{indentfirst}
\usepackage{psfrag}
\usepackage{epsfig}

\renewcommand{\Im}{\mathrm{Im}}
\newcommand{\nbth}{n_b^{\mathrm{th}}}

\newcommand{\xzpf}{x_\text{zpf}}

\usepackage{color} 

\definecolor{darkgreen}{rgb}{0.0, 0.4, 0.26}

\newcommand{\bra}[1]{\ensuremath{\left\langle#1\right|}}
\newcommand{\ket}[1]{\ensuremath{\left|#1\right\rangle}}
\newcommand{\mean}[1]{\ensuremath{\left\langle#1\right\rangle}}

\begin{document}

\title{Molecular optomechanics in the anharmonic regime: from nonclassical mechanical states to mechanical lasing}

\author{Miko\l aj~K.~Schmidt}
\email{mikolaj.schmidt@mq.edu.au}
\affiliation{School of Mathematical and Physical Sciences, Macquarie University, NSW 2109, Australia}

\author{M.~J.~Steel}
\affiliation{School of Mathematical and Physical Sciences, Macquarie University, NSW 2109, Australia}

\begin{abstract}
Cavity optomechanics aims to establish optical control over vibrations of mechanical systems, to heat, cool or to drive them toward coherent, or nonclassical states. This field was recently extended to include molecular optomechanics, which describes the dynamics of THz molecular vibrations coupled to the optical fields of lossy cavities via Raman transitions, and was developed to understand the anomalous amplification of optical phonons in Surface-Enhanced Raman Scattering experiments. But the molecular platform should prove suitable for demonstrating more sophisticated optomechanical effects, including engineering of nonclassical mechanical states, or inducing coherent molecular vibrations. In this work, we propose two pathways towards implementing these effects, enabled or revealed by the strong intrinsic anharmonicities of molecular vibrations. First, to prepare a nonclassical mechanical state, we propose an incoherent analogue of the mechanical blockade, in which the molecular aharmonicity and optical response of hybrid cavities isolate the two lowest-energy vibrational states. Secondly, we show that for a strongly driven optomechanical system, the anharmonicity can effectively suppress the mechanical amplification, shifting and reshaping the onset of coherent mechanical oscillations. Our estimates indicate that both effects should be within reach of the existing implementations of the Surface Enhanced Raman Scattering, opening the pathway towards the coherent and nonclassical effects in molecular optomechanics.
\end{abstract}

\maketitle


\section{Introduction}

Recently, in response to surprising experimental results~\cite{zhu_quantum_2014,zhang_chemical_2013}, it has been suggested that Raman scattering of light from molecules in plasmonic cavities can be cast as an optomechanical process~\cite{roelli_molecular_2016,schmidt_quantum_2016,kamandar_dezfouli_quantum_2017,shlesinger_integrated_2021,lombardi_pulsed_2018,neuman_quantum_2019}, with the molecular vibrations modes playing the role of ultra-high frequency mechanical resonators. This realization brought the vast set of tools developed for canonical cavity optomechanics to the field of Surface- or Tip-Enhanced Raman Scattering (SERS and TERS) research~\cite{le_ru_principles_2009,kneipp_population_1996,itoh_toward_2023,langer_present_2020}. The resulting formalism of \textit{molecular optomechanics} led to new insights into the correlations of the inelastically scattered Raman light~\cite{schmidt_quantum_2016,anderson_two-color_2018}, control over the quantum-mechanical description of the single- and multi-mode plasmonic cavities~\cite{kamandar_dezfouli_quantum_2017,dezfouli_molecular_2019,zhang_addressing_2021,shlesinger_integrated_2021}, or the dynamics of systems with multiple molecules~\cite{zhang_optomechanical_2020}. It also enabled the theoretical proposals~\cite{roelli_molecular_2020}, and experimental demonstrations of new THz detection techniques~\cite{xomalis_detecting_2021,chen_continuous-wave_2021}.

\begin{figure}
    \includegraphics[width=.8\linewidth]{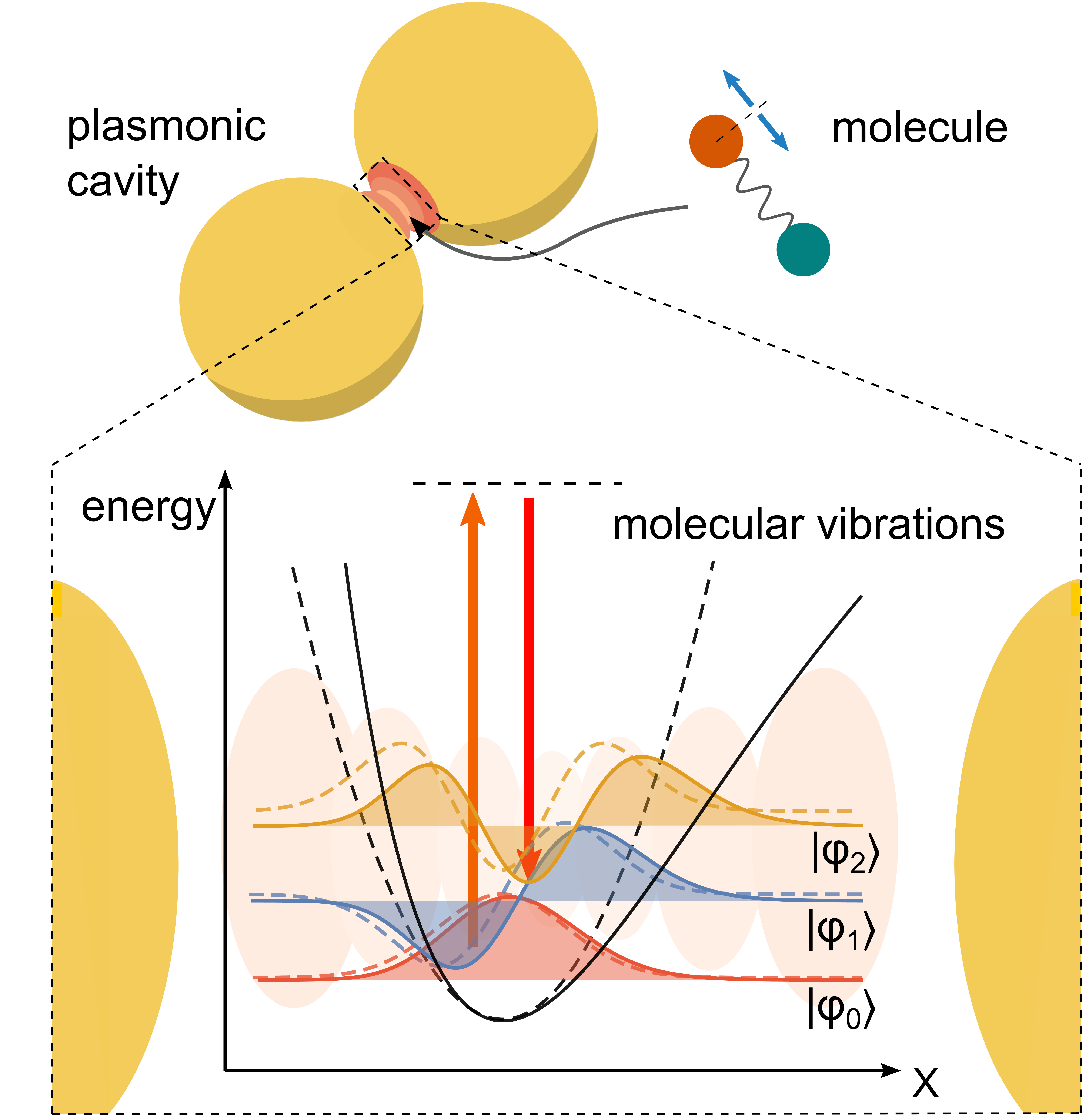}
    \caption{Schematic of the anharmonic molecular optomechanics setup. Molecule, placed in the gap of a plasmonic cavity, experiences off-resonant Raman transitions between the vibrational states of its ground electronic manifold. The anharmonic nature of the potential yields uneven spectrum of the vibrational modes, which can be either used to separate the dynamics of the two lowest-energy states, implementing an acoustic two-level system, or suppress the formation of acoustic lasing transition.}
    \label{fig:schematic}
\end{figure}

Simultaneously, molecular optomechanics stretched the landscape of the conventional cavity optomechanics~\cite{aspelmeyer_cavity_2014} towards the largely unexplored regimes of ultra-high mechanical frequencies characteristic of molecular vibrations, complex (multimode) optical spectrum, and to systems with hundreds, or thousands of homogeneous, and both directly and indirectly coupled mechanical modes. It also brought optomechanics closer to the elusive limit of the strong single-photon coupling~\cite{rabl_photon_2011,nunnenkamp_single-photon_2011}, by confining molecules in ultra-small volume optical cavities~\cite{benz_single-molecule_2016}.

Unlike in the canonical optomechanical systems, the dynamics of the THz molecular vibrations involves only a few, lowest-energy mechanical states, thanks to the combination of low thermal population ($\nbth < 0.1$)~\cite{kneipp_population_1996,le_ru_principles_2009,shalabney_coherent_2015}, large mechanical losses, and small populations of the optical cavities, which render the mechanical amplification mechanism ineffective~\cite{roelli_molecular_2016,schmidt_quantum_2016,benz_single-molecule_2016}. Therefore, in the original formulation of molecular optomechanics~\cite{roelli_molecular_2016,schmidt_quantum_2016}, and follow-up contributions, the molecular vibrations was routinely approximated by a harmonic model of the potential (see the schematic representation in Fig.~\ref{fig:schematic}). However, recent experiments (see e.g. Ref.~\cite{lombardi_pulsed_2018}) are beginning to explore the more efficient amplification mechanisms, setting up questions about the role of the \textit{intrinsic anharmonicity of the mechanical potential}~\cite{witte_femtosecond_2004,morichika_molecular_2019,ventalon_coherent_2004}. To date, these effects were simply neglected, and their potential experimental observations attributed with other effects, such as the bond breaking~\cite{lombardi_pulsed_2018}. At the same time, new types of hybrid plasmonic-dielectric cavities, characterized with small optical mode volumes and sharp spectral features~\cite{shlesinger_integrated_2021,dezfouli_molecular_2019,palstra_hybrid_2019,barreda_hybrid_2022} offer realistic designs of systems which could resolve these anharmonicities.

Therefore, in this work we ask if this anharmonicity can be harnessed for new physics, and how do the predictions of cavity optomechanics hold up in a system with a strong mechanical nonlinearity: Can the anharmonic mechanical potential open up a pathway towards vibrational quantum nonlinearity~\cite{chang_quantum_2014}, and expand the toolbox of nonlinear mechanical phenomena explored to date in optomechanics~\cite{shevchuk_optomechanical_2015,lu_steady-state_2015,sfendla_extreme_2021,shen_mechanical_2022}? Can we use it to engineer nonclassical mechanical states of vibrations~\cite{rips_steady-state_2012} without employing external nonlinear elements~\cite{sanchez_munoz_hybrid_2018,bergholm_optimal_2019}, or can the current experiments induce coherent mechanical lasing~\cite{shishkov_connection_2019,ludwig_optomechanical_2008,jiang_high-frequency_2012} in the presence of the mechanical linearity?

The manuscript is structured as follows: In Section~\ref{sec:formalism} we introduce the formalism, and corrections to the conventional framework of molecular optomechanics, including the redefined coupling mechanism. We then show how this anharmonicity can be harnessed to prepare the molecular vibrations in a nonclassical state (Section~\ref{sec:nonclassical}), and how the anhamornicities modify the mechanical amplification, and the onset of mechanical lasing in molecular systems (Section~\ref{sec:amplification.lasing}).

\section{Formalism}\label{sec:formalism}

In the elementary formulation of molecular optomechanics, we consider a single quantized optical mode, coupled through nonlinear interaction to a single mechanical mode~\cite{roelli_molecular_2016,schmidt_quantum_2016}. The dynamics of this setup is described by the sum of the optical, mechanical, and interaction Hamiltonians:
\begin{equation}
    \hat{H} = \hat{H}_\text{opt} + \hat{H}_M + \hat{H}_{om}.
\end{equation}
The optical mode has resonant frequency $\omega_a$, and is coherently driven with amplitude $\Omega$ and frequency $\omega_l$, so that in the frame rotating with $\omega_l$ we have $\hat{H}_\text{opt} = \hbar (\omega_a-\omega_l) \hat{a}^\dag \hat{a} + \Omega(\hat{a}+\hat{a}^\dag)$. To explicitly consider the harmonic or anharmonic characteristics of molecular vibrations, here we explicitly write the mechanical Hamiltonian as 
\begin{equation}
    \hat{H}_M = \frac{\hat{p}^2}{2m} + V_M(\hat{x}),
\end{equation}
with $V_M$ representing the Morse potential
\begin{equation}
    V_M(x) = D_e \left(1-e^{-\alpha(x-x_0)}\right)^2.
\end{equation}
The eigenfrequency of the $k$th, out of $N$ bound states of $\hat{H}_M$, is
\begin{equation}\label{eq:spectrum}
    \omega_k = \omega_b \left(k+\frac{1}{2}\right) -
\delta \omega_b \left(k+\frac{1}{2}\right)^2,
\end{equation}
with $\delta \omega_b = \hbar \omega_b^2/(4D_e)$, and $N=(\omega_b/\delta \omega_b -1)/1$. 

The optomechanical interaction between molecular system (characterized by the position operator $\hat{x}$) and the optical cavity mode (with electric field operator $\hat{\mathbf{E}}$) is mediated by the Raman dipole, induced in the molecule with Raman polarizability tensor $\mathbf{R}$ by the optical cavity field as
\begin{equation}
    \hat{\mathbf{p}}_R = \mathbf{R}\hat{x}\hat{\mathbf{E}}.
\end{equation}
The explicit connection to the molecular vibrations is given by representing $\hat{x}$ in the basis of the eigenstates $\ket{\phi_k}$ of the mechanical Hamiltonian:
\begin{equation}\label{eq:x}
    \hat{x} = \sum_{kj} x_{k,j} \hat{\sigma}_{k,j} = \underbrace{\sum_{k<j} x_{k,j} \hat{\sigma}_{k,j}}_{\hat{x}^{(-)}} + \underbrace{\sum_{k>j} x_{k,j} \hat{\sigma}_{k,j}}_{\hat{x}^{(+)}}+ \underbrace{\sum_{k} x_{k,k} \hat{\sigma}_{k,k}}_{\hat{x}^{(0)}},    
\end{equation}
where $\hat{\sigma}_{kj}=\ket{\phi_k}\bra{\phi_j}$ denotes the transition operator. The analytical expressions for the matrix elements $x_{kj}=\bra{\phi_k}\hat{x}\ket{\phi_j}$ are given in Appendix~\ref{sec:appendix_position}. Note that in contrast to the harmonic model of optomechanics, the anharmonic potential introduces diagonal components to the $\hat{x}$ operator, represented by $\hat{x}^{(0)}$. The interaction Hamiltonian of the system takes the form~\cite{schmidt_linking_2017,zhang_addressing_2021,kamandar_dezfouli_quantum_2017}: 
\begin{equation}
    \hat{H}_{om} = -\frac{1}{2} \hat{\mathbf{p}}_R \hat{\mathbf{E}} = -\frac{1}{2} \hat{\mathbf{E}}\mathbf{R} \hat{x} \hat{\mathbf{E}} \approx -\mathbf{E}_0(\mathbf{r}) \mathbf{R}\mathbf{E}_0^*(\mathbf{r}) \hat{x} \hat{a}^\dag \hat{a},
\end{equation}
where $\hat{\mathbf{E}}(\mathbf{r}) = \mathbf{E}_0(\mathbf{r})\hat{a}^\dag + \text{h.c.}$, and we carried out the rotating wave approximation to remove the optical mode-squeezing terms ($\propto \hat{a}^2,(\hat{a}^\dag)^2$). 

We note that this framework explicitly assumes an off-resonant nature of the Raman scattering from a single molecule --- see Refs.~\cite{neuman_quantum_2019, shen_optomechanical_2023, shishkov_connection_2019} for the models of the Surface-Enhanced Resonant Raman Scattering (SERRS), and Ref.~\cite{zhang_addressing_2021} for the extension of the off-resonant molecular optomechanics towards multiple molecules.

Since the cavity is coherently driven, and only weakly coupled to the mechanical system, the optical mode fluctuates around a coherent state with amplitude $\mean{\hat{a}} \approx \alpha = \Omega/[-i(\omega_a-\omega_l)-\kappa/2]$, where $\kappa$ is the optical decay rate. We can then linearize the interaction Hamiltonian by separating the coherent part of the optical mode from the fluctuations $\hat{a} = \delta \hat{a}+\mean{\hat{a}} \approx \delta \hat{a}+\alpha$ as:
\begin{equation}
    \hat{H}_{om} = -G_0 (\hat{x}^{(+)}+\hat{x}^{(-)}+\hat{x}^{(0)})(\alpha^* + \delta\hat{a}^\dag)(\alpha + \delta\hat{a}),
\end{equation}
where $G_0 = \mathbf{E}_0(\mathbf{r}_d) \mathbf{R} \mathbf{E}_0^*(\mathbf{r}_d)$, with $\mathbf{r}_d$ denoting the position of the molecule. 

From here, the optomechanical linearization neglects the terms $\propto \delta\hat{a}^\dag\delta\hat{a}$ to write $\hat{H}_{om}\approx \hat{H}_{om}^{(0)} + \hat{H}_{om}^{(\pm)}$ with
\begin{equation}
    \hat{H}_{om}^{(\pm)} = -G_0 \alpha (\hat{x}^{(+)}+\hat{x}^{(-)})(\delta\hat{a}^\dag + \delta\hat{a}) -G_0 \alpha^2 (\hat{x}^{(+)}+\hat{x}^{(-)}),
\end{equation}
where we assumed, without the loss of generality, that $\alpha$ can be made real. The diagonal contributions to the displacement operator $\hat{x}^{(0)}$ define
\begin{equation}\label{eq:Hom.0}
    \hat{H}_{om}^{(0)} =  - G_0 \alpha\hat{x}^{(0)}(\delta\hat{a}^\dag + \delta\hat{a})-G_0 \alpha^2 \hat{x}^{(0)}.
\end{equation}
The second term in $H_{om}^{(0)}$ introduces a shift of the frequency of each vibrational level $\ket{\phi_i}$, mentioned also in Ref.~\cite{jakob_giant_2023}, proportional $\alpha^2 G_0 x_{k,k}$. We thus \textit{dress} the Morse potential eigenfrequencies given in Eq.~\eqref{eq:spectrum} as
\begin{equation}\label{eq:shifts}
    \omega_k \rightarrow \tilde{\omega}_k = \omega_k-\alpha^2 G_0 x_{k,k}.
\end{equation}
For the Morse potential we can find a good approximation for the diagonal matrix elements $x_{k,k}$ (see Appendix~\ref{sec:appendix_position}). 
For reference, we note that the difference between the neighboring eigenfrequencies is
\begin{equation}\label{eq:shifts.diff}
    \tilde{\omega}_k-\tilde{\omega}_{k-1}=\omega_b-2k\delta \omega_b - \alpha^2 G_0 (x_{k,k}-x_{k-1,k-1}),
\end{equation}
and can be approximated using the analytical expressions for the diagonal matrix elements $x_{k,k}$. In particular, as we show in Appendix~\ref{sec:appendix_position}, the shift due to the diagonal term is nearly constant for all $k$.

\subsection{Master equation for the anharmonic molecular vibrations}

From here, we can formulate the effective description of the mechanical state by embracing the quantum noise approach~\cite{marquardt_quantum_2008,clerk_introduction_2010}, treating the driven optical mode as a structured reservoir, and following the evolution of the reduced density matrix $\rho$ of the mechanical system. The coupling to the mechanical mode is then determined by the interaction Hamiltonian~\cite{neuman_quantum_2019,roelli_molecular_2020}, including both $\hat{H}_{om}^{(\pm)}$ and the first term in $\hat{H}_{om}^{(0)}$ in Eq.~\eqref{eq:Hom.0}. In the secular approximation, interaction $\hat{H}_{om}^{(\pm)}$ dictates that the mechanical excitation and decay rates will be given by the spectra of two-time correlators $\mean{[\delta \hat{a}(\tau)]^\dag \delta \hat{a}(0)}$, calculated at the dressed transition frequencies $\tilde{\omega}_k-\tilde{\omega}_j$:
\begin{align}\label{eq:me}
    \dot{\rho} = &-\frac{i}{\hbar}\left[ \sum_{j,k} (\tilde{\omega}_{j}-\tilde{\omega}_k) \hat{\sigma}_{j,k}, \rho \right] \\ \nonumber
    &+ \frac{1}{2} \sum_{k>j}  \frac{g^2\kappa/2}{(\omega_a-\omega_l+\tilde{\omega}_{k}-\tilde{\omega}_j)^2+(\kappa/2)^2} \mathcal{D}\left[\frac{x_{k,j}}{\xzpf}\hat{\sigma}_{k,j} \right] \rho\\ \nonumber
    &+ \frac{1}{2} \sum_{k<j}  \frac{g^2\kappa/2}{(\omega_a-\omega_l+\tilde{\omega}_{k}-\tilde{\omega}_j)^2+(\kappa/2)^2} \mathcal{D}\left[\frac{x_{k,j}}{\xzpf}\hat{\sigma}_{k,j} \right] \rho\\ \nonumber
    &+ \frac{1}{2}\gamma (\nbth + 1)\sum_k \mathcal{D}\left[\frac{x_{k,k+1}}{\xzpf}\hat{\sigma}_{k,k+1}\right] \rho \\ \nonumber
    &+ \frac{1}{2}\gamma \nbth \sum_k\mathcal{D}\left[\frac{x_{k,k-1}}{\xzpf}\hat{\sigma}_{k,k-1} \right] \rho,
\end{align}
where $\mathcal{D}[\hat{O}] \rho = 2\hat{O}\rho \hat{O}^\dag - \hat{O}^\dag\hat{O}\rho- \rho\hat{O}^\dag\hat{O}$ is the GKSL operator, and $g=\alpha G_0 \xzpf$ is the effective optomechanical coupling rate, and $\xzpf=\sqrt{\hbar/(2 m \omega_b)}$ is the zero-point fluctuation of the harmonic oscillator with frequency $\omega_b$. For the more direct comparison with the canonical cavity optomechanics, we normalize the jump operators by the corresponding matrix elements $x_{k,j}/\xzpf$, and separate the first two terms which describe the \textit{Stokes} (mechanical excitation), and \textit{anti-Stokes} (mechanical relaxation) processes, respectively. The remaining two terms describe the effects of the coupling to a thermal bath. For simplicity, we assume that those would be completely described by transitions between neighboring eigenstates, through the constant mechanical decay rate $\gamma$, and the thermal bath population approximated by the Bose-Einstein population $\nbth$ at frequency $\omega_b$. 

Finally, we note that the mechanical system will exhibit an entirely incoherent dynamics, and thus omit the effect of the interaction term $- G_0 \alpha\hat{x}^{(0)}(\delta\hat{a}^\dag + \delta\hat{a})$ in the interaction Hamiltonian $\hat{H}_{om}^{(0)}$ (Eq.~\eqref{eq:Hom.0}), which does not change the mechanical state, and yields an irrelevant, dephasing-like term.

\subsubsection{Multiple optical modes}

Realistic optical systems used in SERS, like the plasmonic nano- and pico-cavities~\cite{benz_single-molecule_2016,lombardi_pulsed_2018,xomalis_detecting_2021,chen_continuous-wave_2021}, or hybrid metallic-dielectric systems~\cite{shlesinger_integrated_2021,dezfouli_molecular_2019}, support multiple overlapping and interacting optical modes, which significantly influence the optomechanical dynamics. In particular, the driving, Stokes and anti-Stokes processes (identified in canonical optomechanics with phonon heating and cooling) can all be mediated via different optical modes. 

An extension of the single-mode model to a multi-mode one presents several difficulties: for example, the incident laser can couple to more than one cavity mode, complicating the definition of coherent amplitude $\alpha$; similarly, the optomechanical coupling parameters $g_0$ needs to be redefined to explicitly account for coupling with modes of the cavity with different field distributions. These difficulties are typically addressed by generalizing the master equations presented above, following the prescriptions from \cite{kamandar_dezfouli_quantum_2017,zhang_addressing_2021,jakob_giant_2023}, which introduce the explicit Stokes $\Gamma_+^{(k)}$ and anti-Stokes $\Gamma_-^{(k)}$ rates, calculated using the Green's function of the system, which completely account for the complex position- and frequency-dependent field distributions of the electric field in the cavity. (see Appendix~\ref{sec:appendix_me} for the definitions and discussion). Adapting this approach to the anharmonic systems, and assuming that the transitions are limited to the neighboring mechanical states, we can rewrite Eq.~\eqref{eq:me} as
\begin{align}\label{eq:me2}
    \dot{\rho} = &-\frac{i}{\hbar}\left[ \sum_{jk} (\tilde{\omega}_{j}-\tilde{\omega}_k) \hat{\sigma}_{jk}, \rho \right] \\ \nonumber
    &+ \frac{1}{2} \sum_{k} \Gamma^{(k)}_+ \mathcal{D}\left[\frac{x_{k+1,k}}{\xzpf}\hat{\sigma}_{k+1,k} \right] \rho\\ \nonumber
    &+ \frac{1}{2} \sum_{k} \Gamma_-^{(k)} \mathcal{D}\left[\frac{x_{k-1,k}}{\xzpf}\hat{\sigma}_{k-1,k} \right] \rho\\ \nonumber
    &+ \frac{1}{2}\gamma (\nbth + 1)\sum_k \mathcal{D}\left[\frac{x_{k,k+1}}{\xzpf}\hat{\sigma}_{k,k+1}\right] \rho \\ \nonumber
    &+ \frac{1}{2}\gamma \nbth \sum_k\mathcal{D}\left[\frac{x_{k,k-1}}{\xzpf}\hat{\sigma}_{k,k-1} \right] \rho,
\end{align}
In particular, for the system with a single optical mode, we can formally write
\begin{equation}\label{eq:def.rates}
    \Gamma^{(k)}_{\pm} = |g_0 \alpha|^2 S_\text{opt,single}(\omega_l - \tilde{\omega}_{k\pm 1} + \tilde{\omega}_{k}),
\end{equation}
where $S_\text{opt,single}(\omega)=\int \textrm{d}\tau\exp(i\omega\tau)\mean{[\delta \hat{a}(\tau)]^\dag \delta \hat{a}(0)}$, simplifying Eq.~\eqref{eq:me2} to Eq.~\eqref{eq:me}. To maintain this intuitive picture and the connection to the canonical optomechanics in the multi-mode case, throughout this work we will assume that the laser selectively couples to only one particular mode of the cavity, and define $|\alpha|^2$ as the \textit{cavity population}. Furthermore, we will fold the entire frequency dependence of the Stokes and anti-Stokes rates into the generalized optical spectrum $S_\text{opt}(\omega)$, while keeping $g_0$ constant. This is a fairly informal step, but it will allow us to develop an intuitive picture of the anharmonic molecular dynamics in multi-mode optical systems.

The master equation in Eq.~\eqref{eq:me2} sets up the dynamics of the system in terms of the diagonal elements of the mechanical density matrix, or the populations of the vibrational states $\rho_{k,k}$, and the total transition rates from the $k$th state, which include the contributions from the unstructured thermal bath 
\begin{equation}\label{eq:rates_thermal}
    \bar{\Gamma}_+^{(k)} = \Gamma_+^{(k)} + \nbth \gamma,~\text{and}~\bar{\Gamma}_-^{(k)} = \Gamma_-^{(k)} + (\nbth+1) \gamma. 
\end{equation}
We list the dynamical equations for these population in Appendix~\ref{sec:appendix_populations} and, in the following sections, investigate their solutions in two cases: weakly pumped, strongly anharmonic system, and strongly pumped system with weaker anharmonicity. 

\section{Nonclassical mechanical states}\label{sec:nonclassical}

To prepare nonclassical states of molecular vibrations in an incoherently driven anharmonic system, we need to suppress its the excitation beyond the two lowest-order states $\left\{\ket{\phi_0},\ket{\phi_1}\right\}$~\cite{chang_quantum_2014} --- that is, form a phonon blockade. Similar schemes have been explored in other branches of physics, most famously in circuit QED, where the Kerr nonlinearity enables the generation of nonclassical states of superconducting circuits~\cite{blais_circuit_2021}. However, that functionality is enabled by the presence of the coherent microwave drive, which induces transitions between specific levels of the anharmonic ladder. In the molecular optomechanics, as well as the canonical cavity optomechanics, all transitions are due to incoherent processes, and so we turn to engineering the rates of these incoherent processes $\Gamma_\pm^{(k)}$ defined above, by structuring the optical spectrum of the system. 

\begin{figure}
    \includegraphics[width=\linewidth]{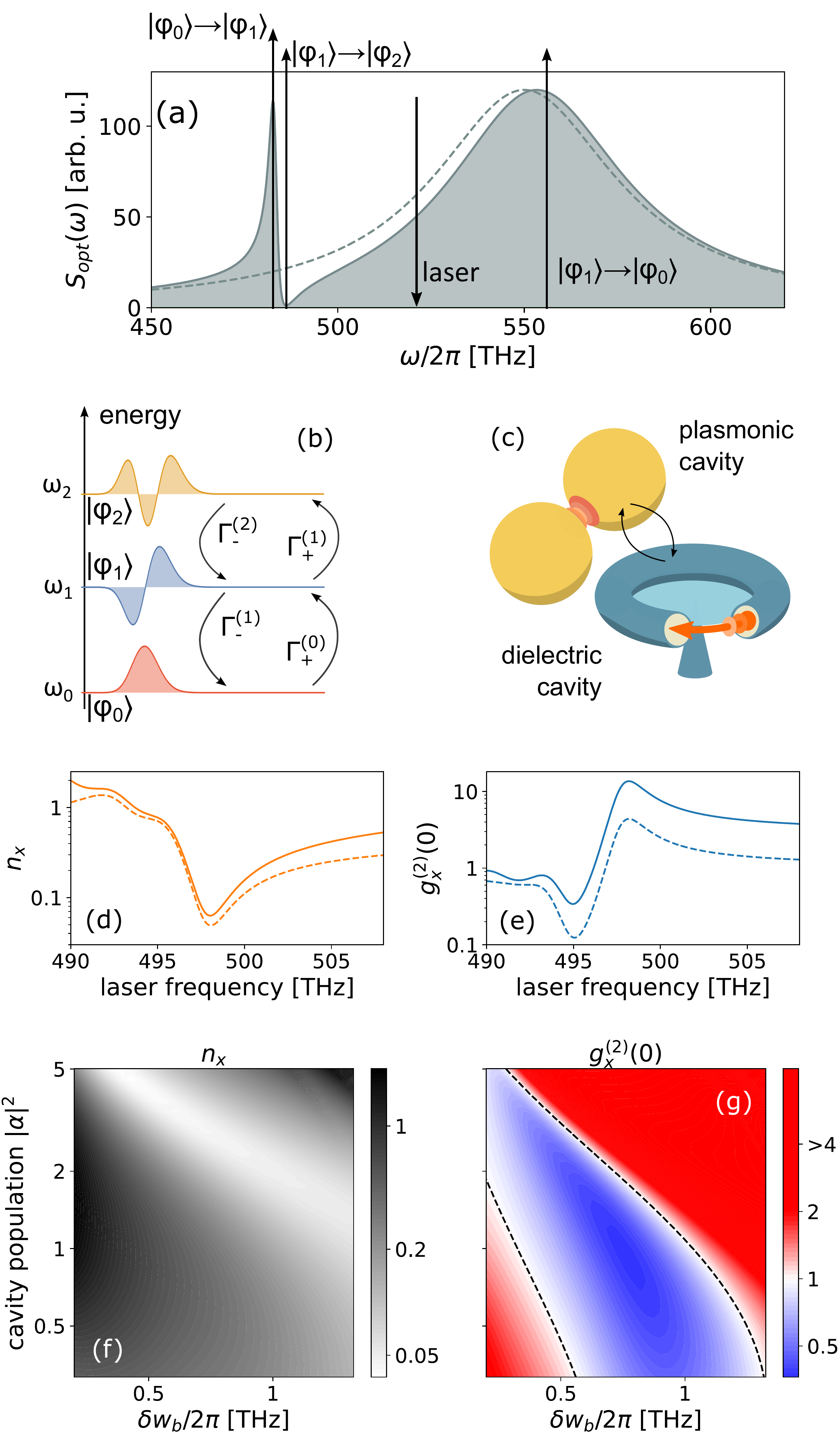}
    \caption{Response of the anharmonic molecular optomechanical setup. (a) Optical spectrum of the hybrid cavity (c) governs the rates of incoherent pumping ($\Gamma_+^{(k)}$) and decay ($\Gamma_-^{(k)}$) (b). For the anharmonic system, Stokes transitions $\ket{\phi_0}\rightarrow \ket{\phi_1}$ and $\ket{\phi_1}\rightarrow \ket{\phi_2}$ occur at slightly different frequencies. (c,e) Mechanical populations (see Eq.~\eqref{eq:phonon_population}) and (d,f) intensity correlations (Eq.~\eqref{eq:correlations}) calculated by solving the rates equations for population (solid lines in (d-e)), and the three lowest-energy states (dashed lines; see text and Eq.~\eqref{eq:g2.approx}), as a function of the incident laser frequency ((c,d), with $\delta \omega_b/2\pi=2$~THz, $g_0|\alpha|/2\pi=4$~THz, or the anharmonicity $\delta \omega_b$ and cavity population $|\alpha|^2$ ((e,f), for $\omega_l/2\pi=501$~THz). Thermal population is set to $\nbth=0.05$.}
    \label{fig:spectra_nonclassical}
\end{figure}

An example of a system with the desirable spectrum is depicted in Fig.~\ref{fig:spectra_nonclassical}(c), where we show a (not to scale) schematic of a hybrid metallic-dielectric cavity, explored in several recent studies \cite{shlesinger_integrated_2021,dezfouli_molecular_2019,palstra_hybrid_2019,barreda_hybrid_2022}. Here, a dimer of gold nanoparticles supports a lossy ($Q\sim 10$) plasmonic mode with dramatically reduced effective mode volumes $\sim 10^{-6} \lambda^3$~\cite{benz_single-molecule_2016}, and coupled to a high-Q dielectric microresonator in the form of toroidal~\cite{shlesinger_integrated_2021}, or nanobeam cavity~\cite{dezfouli_molecular_2019}. The optical response $S_\text{opt}$ of this system, defined in the previous section, is shown in Fig.~\ref{fig:spectra_nonclassical}(a), and exhibits a strong Fano feature due to the off-resonant interaction between the high- and low-Q optical modes. Expressions used to model $S_\text{opt}$ in the presence, and absence of coupling between the modes (depicted with dashed line in  Fig.~\ref{fig:spectra_nonclassical}(a)), and parameters used in this work, are listed in Appendix~\ref{sec:spectra}.

In an idealized scheme, we realize the incoherent blockade by tuning the first Stokes frequency $\omega_l-(\tilde{\omega}_1-\tilde{\omega}_0)$ to matche the peak of the Fano feature, so that state $\ket{\phi_1}$ can be efficiently populated from $\ket{\phi_0}$; additionally, if the second Stokes frequency $\omega_l-(\tilde{\omega}_2-\tilde{\omega}_1)$ matches the dip in the optical spectrum, transitions to the $\ket{\phi_2}$ state become suppressed. This incoherent blockade should be therefore governed by the contrast between the optical spectra calculated at $S_\text{opt}(\omega_l-\tilde{\omega}_{1}+\tilde{\omega}_0)$ and $S_\text{opt}(\omega_l-\tilde{\omega}_{2}+\tilde{\omega}_1)$. A similar scheme, involving multiple high-$Q$ optical modes for controlling the state of a MHz mechanical oscillator with Kerr nonlinearity, was proposed by Rips \textit{et al.}~\cite{rips_steady-state_2012}.

To characterize the population of the mechanical states, and its non-classical statistics, in a manner most resembling the usual second quantization language of populations and statistics of the harmonic system, in Appendix~\ref{sec:app_readout} we employ the position operator $\hat{x}$ as the key observable, used to define the steady-state \textit{mechanical populations} and \textit{intensity correlations}
\begin{equation}\label{eq:phonon_population}
    n_x = \xzpf^{-2}\mean{\hat{x}^{(-)}\hat{x}^{(+)}},
\end{equation}
\begin{equation}\label{eq:correlations}
    g^{(2)}_x(\tau)=\frac{\mean{\hat{x}^{(-)}(0) \hat{x}^{(-)}(\tau) \hat{x}^{(+)}(\tau) \hat{x}^{(+)}(0)}}{\mean{\hat{x}^{(-)}\hat{x}^{(+)}}^2}.
\end{equation}
We note that as the system becomes anharmonic, and we turn away from the second quantization framework, $n_x$ losses its exact definition as the phonon population. Nevertheless, we embrace this language for simplicity, and a direct mapping to the harmonic setup. Furthermore, if we measured the direct IR emission from the transitions between the mechanical states, these magnitudes would characterize the intensity, and statistics of the emitted IR light. Further details about the calculations of $n_x$ and $g^{(2)}_x(\tau)$ are listed in Appendix~\ref{sec:appendix_populations}.

In Fig.~\ref{fig:spectra_nonclassical}(d,e) we plot with solid lines the populations $n_x$ and intensity correlations $g^{(2)}_x(0)$, as a function of the laser frequency $\omega_l$. The former are visibly suppressed when laser is chosen to have its first Stokes frequency $\omega_l-(\tilde{\omega}_{1}-\tilde{\omega}_0$) match the dip in the optical cavity spectrum, around 498~GHz. Conversely, when the second Stokes transition at $\omega_l-(\tilde{\omega}_{2}+\tilde{\omega}_1$) matches the dip in $S_\text{opt}$ for the laser around 495~THz, we approach the blockade condition described above, the system demonstrates sub-Poissonian statistics $g^{(2)}_x(0)<1$.  Away from these features, the system acquires the thermal statistics $g^{(2)}_x(0)\sim 2$. In Fig.~\ref{fig:spectra_nonclassical}(f,g) we show the same magnitudes, calculated as a function of nonlinearity $\delta \omega_b$, and coherent cavity population $|\alpha|^2$. Here again the statistics diverges from the thermal one towards sub-Poissonian (denoted as a blue region), when the second Stokes transition frequency is tuned to the dip of the Fano feature in the optical spectrum $S_\text{opt}$. Since that anti-Stokes frequency explicitly depends on $|\alpha|^2$ due to the dressing by the coherent field (see Eq.~\eqref{eq:shifts.diff}), the region of sub-Poissonian statistics is largely diagonal. In Appendix~\ref{app:laser.freq} we discuss how this region of sub-Poissonian statistics changes with the laser frequency.


To gain analytical insight into these effects, we consider the analytical solution to the coupled equations for the populations $\rho_{k,k}$ of the three lowest-energy states (see derivation in Appendix~\ref{subsec:antibunching_app}), approximate the numerator and denominator in the definition of $g^{(2)}_x(0)$ (Eq.~\eqref{eq:correlations}) by $2\rho_{2,2}$ and $\rho_{1,1}$, respectively, to find
\begin{align}\label{eq:g2.approx}
    g^{(2)}_x(0)&\approx 2\frac{\bar{\Gamma}_+^{(1)}\left[\bar{\Gamma}_-^{(1)}\bar{\Gamma}_-^{(2)}+\bar{\Gamma}_+^{(0)}\left(\bar{\Gamma}_-^{(2)}+\bar{\Gamma}_+^{(1)}\right)\right]}{\bar{\Gamma}_+^{(0)}\left(\bar{\Gamma}_-^{(2)}+2\bar{\Gamma}_+^{(1)}\right)^2}\\
    &\approx 2\frac{\bar{\Gamma}_+^{(1)}}{\bar{\Gamma}_+^{(0)}}\left(1+\frac{\bar{\Gamma}_+^{(0)}}{\bar{\Gamma}_-}\right).
\end{align}
We plot the function given in the first line in Fig.~\ref{fig:spectra_nonclassical}(e) with the dashed line, finding a qualitative agreement of the spectral range corresponding to the sub-Poissonian statistics with the full calculations (solid line). We have verified that the discrepancy is due to the truncation to the three states.

The second line represents a far more crude approximation, where we assume that the anti-Stokes transitions rates $\bar{\Gamma}_-^{(k)}$ are largely independent of $k$, and far larger than the second anti-Stokes rate $\bar{\Gamma}_+^{(1)}$. The first fraction in this expression directly characterizes the contrast in the anti-Stokes due to the Fano feature of the optical cavity. 

As we discuss in more detail in Appendix~\ref{app:laser.freq}, we can further suppress the intensity correlations by increasing the relative role of the optomechanical feedback over the thermal pumping. This can be achieved by employing a larger intensity of the optical driving, although one needs to account for the $|\alpha|^2$ dependence of the dressed mechanical frequencies (Eq.~\eqref{eq:shifts.diff}), to ensure that the transitions $\ket{\phi_0}\rightarrow \ket{\phi_1}$ and $\ket{\phi_1}\rightarrow \ket{\phi_2}$ match the peak, and the trough of the optical spectrum. One could also explore hybrid resonators featuring larger contrasts of the peak and troughs of the Fano resonance.

In numerical modelling, we assumed a single molecule exhibiting optomechanical coupling $g_0/2\pi=2$~THz, consistent with the values reported for picocavities in Ref.~\cite{benz_single-molecule_2016}, and below the estimates for coupling with a small ensemble of about 100 molecules in nanocavities in Ref.~\cite{lombardi_pulsed_2018}. The non-classical state of vibrations in Fig.~\ref{fig:spectra_nonclassical}(g) was reported for aharmonicities as small as $\delta \omega_b/2\pi=0.1$~THz, which yields the anharmonicity parameter $\xi=\delta \omega_b/\omega_b\approx 5\times 10^{-3}$ --- arguably large, but reportedly accessible with molecular systems investigated in the context of SERS~\cite{jakob_giant_2023}. We also choose to plot the results against the population of the driven optical (plasmonic) mode, rather than the input laser intensity. The low populations explored here are typical for a strongly driven, lossy plasmonic cavities~\cite{benz_single-molecule_2016,lombardi_pulsed_2018}, and should offer good approximation to the characteristics of hybrid systems under the assumption that the laser predominantly drives the plasmonic mode.

Finally, we note that since our scheme is based around harnessing the changes to the optical spectrum between the two Stokes transitions, it does require the mechanical system to exhibit a strong nonlinearity $\delta \omega_b$ to resolve the features of the optical spectrum. However, it does not require the mechanical system to operate in the conventional phonon blockade regime $\delta \omega_b>\gamma$. 

\section{Mechanical amplification and lasing}\label{sec:amplification.lasing}

Anharmonicity of molecular vibrations should also have a strong effect on the opposite regime of molecular optomechanics, where the system is strongly optically driven, in the effort to amplify the mechanical mode, and boost the intensity of Stokes emission~\cite{kneipp_population_1996,langer_present_2020}. In this scenario, the amplification is likely to be suppressed until it saturates the ladder of bound states of the vibrations, or the non-resonant model of Raman scattering breaks down. We explore these effects in Section~\ref{sec:amplification}. 

Moreover, beyond the amplification regime, canonical optomechanical systems exhibit a transition to mechanical \textit{lasing} or dynamical instability, where the mechanical component exhibits coherent oscillations~\cite{ludwig_optomechanical_2008,shishkov_connection_2019}. In Section~\ref{sec:instability} we explore similar effects in the context of molecular optomechanics, analyzing the impact of anharmonicity on the threshold, amplitude, and the trajectory of these oscillations.

\subsection{Amplification}\label{sec:amplification}

\begin{figure}
    \includegraphics[width=\linewidth]{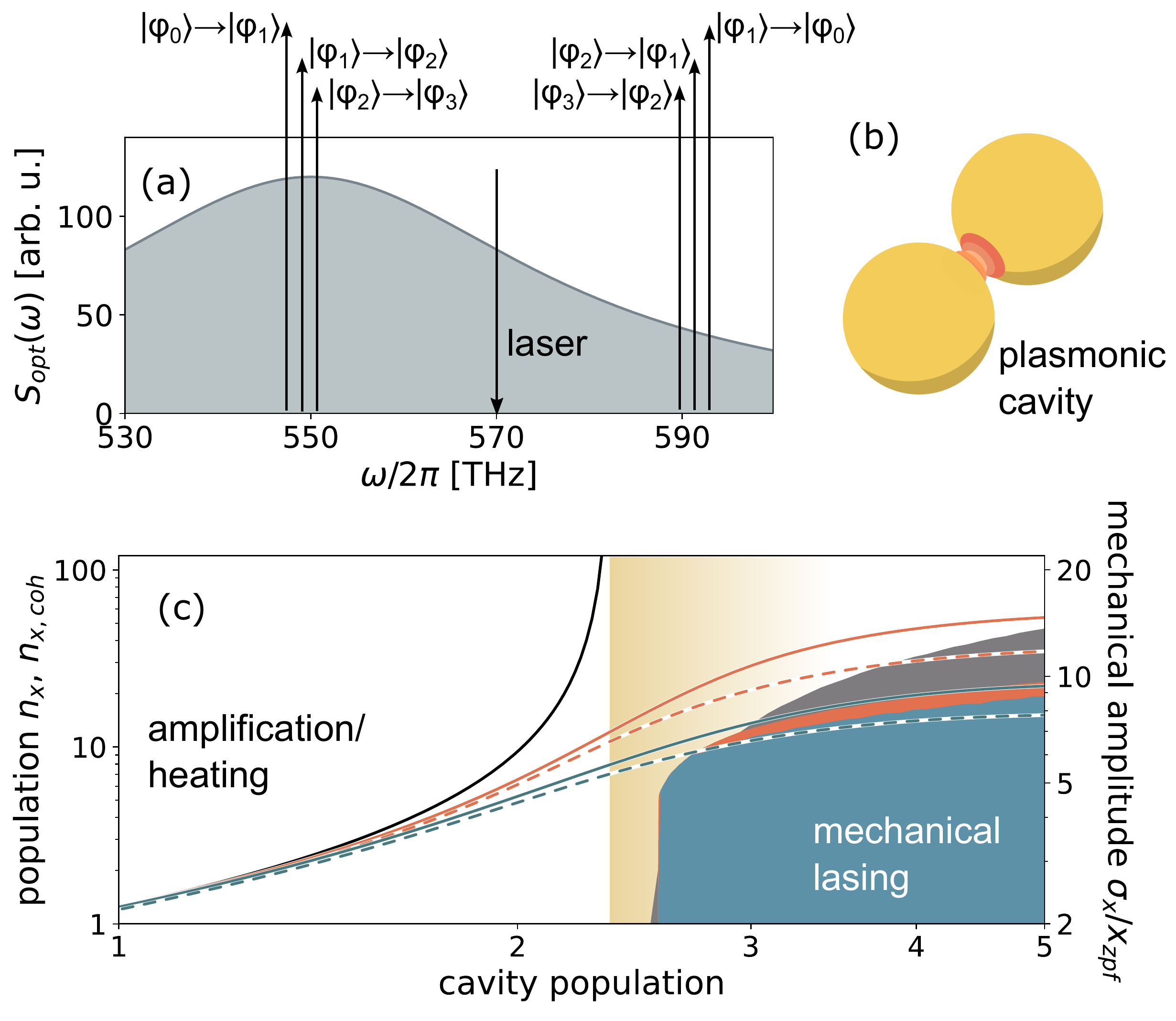}
    \caption{Amplification and mechanical instability in ahramonic molecular optomechanics. (a) Stokes and anti-Stokes transition rates ($\Gamma_+^{(k)}$ and $\Gamma_-^{(k)}$), between every two neighboring mechanical levels depend on the spectra of the single-mode optical (plasmonic) cavity. The shifts are exaggerated for clarity. (b) Phonon populations $n_x$ calculated using the full model (solid lines), analytical solution to the Eq.~\eqref{eq:rate.equation} (dashed lines) and standard deviations of the mechanical oscillations $\sigma_x$ (filled areas) as a function of the optical cavity population. Black line denotes results for the harmonic system, while the red, and teal lines correspond to increasing anharmonicity $\delta \omega_b/2\pi=0.1$, and~0.2~THz.}
    \label{fig:results_lasing}
\end{figure}

Using the formulation developed in Section~\ref{sec:nonclassical}, we can readily explore the mechanical amplification by analyzing the steady-state populations mechanical $n_x$ as a function of the input pump power, or the cavity population. To this end, we consider a simpler optical setup, with a single optical mode ($S_\text{opt}=S_\text{opt,single}$) being driven with a blue-detuned laser shown schematically in  Fig.~\ref{fig:results_lasing}(a), to promote the Stokes emission, and mechanical amplification. To describe the anharmonicity, we approximate the rates of the non-thermal Stokes and anti-Stokes processes by expanding the cavity spectrum $S_\text{opt}(\omega)$ as 
\begin{align}
    &S_\text{opt}(\omega_l- \tilde{\omega}_{k+1} + \tilde{\omega}_{k}) \\ 
    &\approx S_\text{opt}\left(\omega_l - \omega_b + \alpha^2 g_0 \sqrt{\frac{\omega_b}{\delta\omega_b}}\frac{3+2/N}{2N+1}\right) + \eta_+ (k+1) 2\delta \omega_b,
\end{align}
and
\begin{align}
    &S_\text{opt}(\omega_l- \tilde{\omega}_{k-1} + \tilde{\omega}_{k}) \\ 
    &\approx S_\text{opt}\left(\omega_l + \omega_b - \alpha^2 g_0 \sqrt{\frac{\omega_b}{\delta\omega_b}}\frac{3+2/N}{2N+1}\right) - \eta_- k 2\delta \omega_b,
\end{align}
where we used Eq.~\eqref{eq:shifts.diff}, and the explicit form of the diagonal matrix elements (Eq.~\eqref{eq:x.diag}). Parameters $\eta_\pm$ are defined as the derivatives of the optical spectra near the frequencies given as arguments of the optical spectra in second lines of the above equations. Per the definition of the Raman transition rates (see Eq.~\eqref{eq:def.rates}), we can approximate them using the above expansion as:
\begin{align}\label{eq:Gamma.plus.minus.def}
    &\Gamma_+^{(k)} \approx \Gamma_+ + \eta_+ (k+1) g^2 2 \delta w_b,\\
    &\Gamma_-^{(k)} \approx \Gamma_- - \eta_- k g^2 2 \delta w_b,
\end{align}
where $\Gamma_+ = g^2 S_\text{opt}[\omega_l - \omega_b + \alpha^2 g_0 \sqrt{\omega_b/\delta\omega_b}(3+2/N)/(2N+1)]$ and $\Gamma_- = g^2 S_\text{opt}[\omega_l + \omega_b - \alpha^2 g_0 \sqrt{\omega_b/\delta\omega_b}(3+2/N)/(2N+1)]$. The exemplary optical spectrum with exaggerated anharmonic shifts, and the corresponding rates, is schematically depicted in Fig.~\ref{fig:results_lasing}(a).

Using this formulation, in Appendix~\ref{sec:corrected.rate.equations} we show that the rate equation for the mechanical population $n_x$ is thus modified from its conventional, linear form, into
\begin{align}\label{eq:rate.equation}
    \frac{\textrm{d}}{\textrm{d}t} n_x
    =-&n_x\left[\left(\gamma+\Gamma_--\Gamma_+ \right) \underbrace{- g^2 2\delta \omega_b \left(\eta_-+5\eta_+\right)}_{\text{linear damping}}\right] \nonumber \\ 
    +&\underbrace{4n_x^2 g^2 \delta \omega_b \left(\eta_-+2\eta_+\right)}_{\text{quadratic damping}} \nonumber \\
    +&\Gamma_+ + \gamma \nbth + \underbrace{\eta_+ g^2 2 \delta \omega_b}_{\text{constant damping}}.
\end{align}
In a single-mode optical cavity, the mechanical lasing setup would require driving on the blue side of the optical resonance (see Fig.~\ref{fig:results_lasing}(a)), in which case the slope of spectrum $S_\text{opt}$ at $\omega_l+ \omega_b$ should be negative ($\eta_-<0$). Thus, we can interpret the additional terms in the rate equation as damping with the linear and quadratic dependence on the mechanical population $n_x$, and a constant term; all are also dependent on the driving intensity, through the coherent cavity population $|\alpha|^2$ in $g^2$, and the frequencies at which derivatives $\eta_\pm$ are calculated.

In Fig.~\ref{fig:results_lasing}(c) we show the steady-state mechanical population $n_x$ as a function of the cavity population for several values of $\delta \omega_b$. Dashed lines denote the $n_x$ derived from Eq.~\eqref{eq:rate.equation}, by setting its LHS to 0, and solving the resulting quadratic equation for $n_x$. Solid lines are calculated by solving the complete set of rate equation, as described in Appendix~\ref{sec:appendix_populations}. For reference, we include the result obtained with the harmonic oscillator (dashed black line), which clearly demonstrates divergence, signalling the conventional threshold of the mechanical lasing.

From Fig.~\ref{fig:results_lasing}, we can derive several observations: 
(i) Even for the smallest anharmonicity (red lines, $\delta \omega_b/2\pi=0.1$~THz) the mechanical amplification is significantly suppressed, compared to the harmonic case; (ii) $n_x$ exhibits a limited superlinear dependence on the cavity population and thus driving intensity; for the stronger nonlinearity (teal lines, $\delta \omega_b/2\pi=0.2$~THz), $n_x$ is either linear or supralinear with $|\alpha|^2$, in a significant deviation from the harmonic optomechanical models, (iii) for neither anharmonic system does the population exhibit a clear mechanical lasing threshold. 

Despite the anharmonicity suppressing the heating of the vibrations, they appear to build up a substantial population $>10$. While this is still far below the limit of $N$ bound states, the Raman transitions between the higher-energy states  are likely to couple to the electronic excited states of the molecule, in a resonant Raman fashion. This effect is naturally absent in the canonical cavity optomechanics.

\subsection{Dynamical instability}\label{sec:instability}

The linearized coupling theory of optomechanics predicts a natural limit to the mechanical amplification, when the system reaches a threshold of the dynamical instability (or phonon lasing), at which point the incoherent phonon population should diverge. In reality near that threshold the linearized formulation fails, the phonon population remains finite, and the mechanical mode begins to exhibit coherent oscillations. These oscillations grow quickly until the system reaches a steady state, with their amplitude dependent on the optical driving strength and detuning from the optical cavity, and the optomechanical coupling~\cite{lorch_sub-poissonian_2015,ludwig_optomechanical_2008,jiang_high-frequency_2012}.

Here, we ask if an optomechanical system with an anharmonic, Morse potential, would similarly exhibit steady-state mechanical solutions near the mechanical lasing threshold. While the complete mapping of this effects --- possible in the harmonic case --- goes beyond the scope of this work, we explore it in the range of parameters relevant for molecular optomechanics.

\subsubsection{Harmonic potential}

The classical trajectories for the optomechanical system can be identified from the dynamical equations for the optical amplitude $\alpha$, and the position and momentum of the mechanical oscillator $x$ and $p$:
\begin{subequations}\label{eq:trajectories.harmonic}
    \begin{equation}
        \frac{\mathrm{d}}{\mathrm{d}t} \alpha = -i\Delta \alpha - i\frac{g_0}{\xzpf}x\alpha - i\Omega - \frac{\kappa}{2} \alpha,\label{eq:trajectories.harmonic.a}
    \end{equation}
    \begin{equation}
        \frac{\mathrm{d}}{\mathrm{d}t} x = 2 \frac{\omega_b \xzpf^2}{\hbar}p - \frac{\gamma}{2} x,\label{eq:trajectories.harmonic.x}    
    \end{equation}
    \begin{equation}
        \frac{\mathrm{d}}{\mathrm{d}t} p = -\frac{\hbar g_0}{\xzpf} |\alpha|^2 - \frac{\hbar \omega_b}{2\xzpf^2}x-  \frac{\gamma}{2} p.\label{eq:trajectories.harmonic.p}
    \end{equation}
\end{subequations}
In the mechanical lasing regime, the mechanical trajectory settles into oscillations $x(t)=x_0 + A\cos(\omega_b t)$, and thus the nonvanishing amplitude $A$ is typically used to characterize both the onset, and magnitude of the oscillations \cite{ludwig_optomechanical_2008,aspelmeyer_cavity_2014,bakemeier_route_2015}. 

We reproduce that result numerically, by rewriting the above coupled equations into a form more suitable for numerical simulations (see Appendix~\ref{app:trajectories} and Ref.~\cite{bakemeier_route_2015}), and solve them with initial conditions $\alpha(0)=x(0) = p(0) = 0$, until the mechanical mode settles into steady-state oscillations. We then characterize these oscillations by their standard deviation $\sigma_x = \sqrt{\mean{(x(t)-\mean{x}_t)^2}_t}$, with the temporal average $\mean{...}_t$ calculated in the steady-state (and nominally equivalent to $A$). Since these oscillations represent coherent dynamics, we introduce the equivalent coherent mechanical population defined as $n_{x,\text{coh}} = (\sigma_x/\xzpf)^2/4$. The normalized deviation $\sigma_x/\xzpf$, and the corresponding $n_{x,\text{coh}}$ for the harmonic system are denoted in Fig.~\ref{fig:results_lasing}(c) as a filled gray area. 

\subsubsection{Anharmonic oscillator}

For the anharmonic oscillator, we can write down similar equations for the amplitude of the optical mode, and coordinates of the mechanical oscillator; the only difference will be the change to the term in Eq.~\eqref{eq:trajectories.harmonic.p} which describes the force $F=-V'(x)$ derived from the harmonic potential ($- x{\hbar \omega_b}/({2\xzpf^2})$) to the corresponding force associated with the Morse potential ($F=-V_M'(x) = - 2D_e a [1-\exp(-ax)]\exp(-ax)$):
\begin{equation}
    \frac{\mathrm{d}}{\mathrm{d}t} p = -\frac{\hbar g_0}{\xzpf} |\alpha|^2 - 2D_e a \left(1-e^{-ax}\right)e^{-ax}-  \frac{\gamma}{2} p.\label{eq:trajectories.anharmonic.p}
\end{equation}
Unlike for the harmonic oscillators, we do not have an analytical solution for the amplitude of the oscillations in the anharmonic system. Nevertheless, we can again characterize them by the standard deviation $\sigma_x/\xzpf$ calculated in the steady state, and the corresponding $n_{x,\text{coh}}$. We depict these results in Fig.~\ref{fig:results_lasing}(c) as the filled areas corresponding to the anharmonic systems with $\delta \omega_b/2\pi=0.1$ (red area), and~0.2~THz (teal area).

These results indicate that mechanical lasing should be possible even within the very limited space of the very few bound states of the mechanical system. For example, for $\delta \omega_b/2\pi=0.2$~THz (teal area), the Morse potential supports about $N=50$ states, and the coherent oscillation have the equivalent coherent population $n_{x,\text{coh}} \gtrsim 5$, comparable to the incoherent populations $n_x$. Much like in Raman or Brillouin lasers, this should lead to observable narrowing of the Raman lines~\cite{grudinin_phonon_2010,kuang_nonlinear_2023}, should such measurements be possible. In Appendix~\ref{app:trajectories} we show how the thresholds, and amplitudes of these coherent oscillations change with anharmonicity, finding that the anharmonicity does not introduce a qualitative change to the amplitude of mechanical oscillations, except for a more pronounced and shifted threshold behaviour, and lowered amplitudes.

\section{Conclusion}\label{sec:conclusion}

In this work, we show that by the framework of molecular optomechanics offers pathways towards genuine quantum engineering of molecular vibrations, far beyond the typical applications to mechanical amplification or Surface-Enhanced Raman Scattering. 

In particular, by embracing the anharmonicity of the molecular vibrations, and engineering the optical spectrum of hybrid cavities, we show how mechanical systems can be driven into the weakly populated, nonclassical states. We present a complete theoretical framework for designing and describing such systems, and formal connection that can serve to characterize the nonclassical statistics of the emitted THz photons. 

In the opposite regime of strong driving, we show that the current experimental setups should be capable of inducing coherent oscillations, or mechanical lasing, of the molecular vibrations. This effect should lead to an observable narrowing of the Raman scattering lines, and further the mapping between molecular, and canonical optomechanics. Furthermore, we show that even weak anharmonicities can dramatically change the optical response of the system, reshaping the dependence of the Stokes intensity on driving power.

These findings should significantly expand the toolbox and the impact of the formalism of molecular optomechanics, towards applications in quantum sensing and microscopy, and call for revisiting of the reported experimental results, to analyze the possible impact of the anharmonicities.

\begin{acknowledgments}
M.K.S. acknowledges support from the Macquarie University Research Fellowship scheme (MQRF0001036) and the Australian Research Council Discovery Early Career Researcher Award DE220101272, and fruitful discussions with J. Aizpurua and R. Esteban.
\end{acknowledgments}

\appendix

\section{Eigenstates of the harmonic and Morse potentials}
\label{sec:appendix_position}

To draw a complete picture of the correspondence between the canonical cavity optomechanics, and anharmonic molecular optomechanics, we first consider the complete expressions for the spectra and spatial characteristics of the eigenstates in either case. 

Schr\"odinger equation with harmonic potential:
\begin{equation}
    V_H (r) = \frac{1}{2} kr^2,
\end{equation}
has eigenfrequencies
\begin{equation}
    E_n = \hbar \omega_0 \left(n+\frac{1}{2}\right), 
\end{equation}
where $\omega_0 = \sqrt{k/m}$, and the matrix elements of the position operator $\hat{x}$ are $\hat{x}^H=\xzpf(\hat{b}^\dag+\hat{b})$, with zero-point fluctuation $\xzpf = \sqrt{\hbar/(2m\omega_b)}$: 
\begin{align}
    &x^H_{n-1,n}=\bra{n-1}\hat{x}\ket{n} = \xzpf\sqrt{n},\\
    &x^H_{n,n-1}=\bra{n}\hat{x}\ket{n-1} = \xzpf\sqrt{n},
\end{align}
and 0 for all other elements.

Schr\"odinger equation with Morse potential:
\begin{equation}
    V_M (r) = D_e \left(1-e^{-ar}\right)^2,
\end{equation}
has eigenstates denoted as $\ket{\phi_n}$ and corresponding eigenfrequencies
\begin{equation}
    \omega_{k} = \omega_b \left(k+\frac{1}{2}\right) - \frac{\hbar \omega_b^2}{4D_e}\left(k+\frac{1}{2}\right)^2.
\end{equation}
In the Dunham expansion we have $Y_{1,0}=\hbar \omega_b$, and $Y_{2,0}=-(\hbar \omega_b)^2/(4D_e)$, and $\delta \omega_b=\hbar \omega_b^2/(4D_e)$.



The matrix elements for the eigenstates of the Morse potential, given in Ref.~\cite{lima_matrix_2005}, and normalized by the zero-point fluctuation, are
\begin{align}
    \frac{x_{n,m}}{\xzpf} = &\bra{\phi_n}\frac{\hat{x}}{\xzpf}\ket{\phi_m}  \\ 
    = &\sqrt{\frac{\omega_b}{\delta \omega_b}}(-1)^{m-n+1}\frac{2}{(n-m)(2N-n-m)} \nonumber \\
    &\times \sqrt{(N-n)(N-m)\frac{\Gamma(2N-n+1)n!}{\Gamma(2N-m+1)m!}},\nonumber
\end{align}
for $n>m$, with $N=(\omega_b/\delta \omega_b-1)/2$, and using the Gamma function $\Gamma$.
We note that these elements are symmetric $x_{n,m}=x_{m,n}$.

In general, we find that these elements do not vanish for non-neighboring states ($n-m=\pm 1$), but drop-off quickly with $|n-m|$. For the neighboring states, we can simplify the above formulation by using the recursive property of the $\Gamma$ function $\Gamma(z+1)=z\Gamma(z)$. In particular, for $m=n+1$, we find
\begin{align}
    \frac{x_{m-1,m}}{\xzpf} = &-\sqrt{\frac{\omega_b}{\delta \omega_b}\frac{(N-m+1)(N-m)m}{2N-m+1}}\frac{2}{2N-2m+1}.\nonumber
\end{align}    

For $n=m$ we find
\begin{align}\label{eq:x.diag}
    \frac{x_{n,n}}{\xzpf} = \bra{\phi_n}\frac{\hat{x}}{\xzpf}&\ket{\phi_n}  \\ 
    = \sqrt{\frac{\omega_b}{\delta \omega_b}}[&\ln(2N+1) + \Psi(2 N-n+1) \nonumber \\ 
    &- \Psi(2N-2n+1) - \Psi(2N-2n)],\nonumber
\end{align}
with the digamma function $\Psi$. In the limit of weak anharmonicity and for low-order states $N\gg n$, we can use its asymptotic property $\Psi(z)\approx \ln(z)-1/(2z)$ to approximate the diagonal terms $x_{n,n}$ as
\begin{align}\label{eq:xnn.approx}
    \frac{x_{n,n}}{\xzpf} \approx \sqrt{\frac{\omega_b}{\delta \omega_b}}\left[2\ln\left(\frac{N+1/2}{N}\right)+\left(3+\frac{2}{N}\right)\frac{n}{2N+1}\right].
\end{align}

\subsection{Shifts of Raman spectra due to the diagonal terms}
Dropping the small, $n$-independent constant term in Eq.~\ref{eq:xnn.approx}, we can use it to approximate the Stokes transition frequency (see Eq.~\eqref{eq:shifts.diff}) from state $\ket{\phi_k}$ to $\ket{\phi_{k+1}}$ as 
\begin{equation}\label{eq:Stokes.freq.approx}
    \tilde{\omega}_{k+1} - \tilde{\omega}_{k}\approx \omega_b - \alpha^2 g_0 \sqrt{\frac{\omega_b}{\delta \omega_b}}\frac{3+2/N}{2N+1} - (k+1)2\delta \omega_b,
\end{equation}
and the anti-Stokes transition frequency from state $\ket{\phi_k}$ to $\ket{\phi_{k-1}}$ as 
\begin{equation}\label{eq:aStokes.freq.approx}
    \tilde{\omega}_{k-1} - \tilde{\omega}_{k}\approx -\omega_b + \alpha^2 g_0 \sqrt{\frac{\omega_b}{\delta \omega_b}}\frac{3+2/N}{2N+1} + k 2\delta \omega_b.
\end{equation}


\section{Master equation in a system with multiple optical modes}
\label{sec:appendix_me}

As discussed in the main text, we assume here that the laser selectively drives a particular optical mode, denoted by $i$, and characterized with a normalized mode distribution $\mathbf{E}_{i, 0}$. As this mode is populated with a coherent state with amplitude $\alpha$, its electric field is given by $\mathbf{E}(\omega_l,\mathbf{r})=\alpha\mathbf{E}_{i, 0}(\omega_l,\mathbf{r})$. From here, we follow the prescription by Zhang~\textit{et al.}~\cite{zhang_addressing_2021}, where the authors introduce transitions rates between the \textit{neighboring} mechanical states $\ket{\phi_k}$ and $\ket{\phi_{k\pm 1}}$, denoted as $\Gamma^{(k)}_{\pm}$, and given by
\begin{align}\label{eq:rate.def.app}
    \Gamma^{(k)}_{\pm} = &\frac{1}{2\hbar \varepsilon_0 c^2} \left[\omega_l \mp (\tilde{\omega}_{k\pm 1} - \tilde{\omega}_{k})\right]^2 \\
    &\times \mathbf{p}^*\cdot \Im\,\mathbf{G}[\mathbf{r}_m,\mathbf{r}_m,\omega_l \mp (\tilde{\omega}_{k\pm 1} - \tilde{\omega}_{k})]\cdot \mathbf{p}, \nonumber
\end{align}
where $\mathbf{G}$ is the electromagnetic Green's functions of the system decoupled from the molecule, $\varepsilon_0$ denotes the vacuum permittivity, and $c$ the speed of light. Finally $\mathbf{p}$ represents the Raman dipole induced by the electric field of the optical mode driven by the laser:
\begin{equation}\label{eq:Raman.dipole}
    \mathbf{p}=\xzpf \mathbf{R}{\mathbf{E}}(\mathbf{r}_d,\omega_l)\approx \xzpf \alpha \mathbf{R}\mathbf{E}_{i,0}(\mathbf{r}_d,\omega_l).
\end{equation}
Unfortunately, in multi-mode systems, the definition or rates given in Eq.~\eqref{eq:rate.def.app} cannot be directly transformed to that inherited from the single-mode setup, and expressed in Eq.~\eqref{eq:def.rates} (see the Supplementary Materials of Ref.~\cite{jakob_giant_2023} for a discussion of this issue in single-cavity systems). That is because the latter model assumes that for any frequency the {field distribution} (and thus the interaction with the molecule) of the sole optical mode will be identical, and only rescaled by a scalar representing the lorentzian spectrum of that optical mode. Conversely, the multi-cavity setup exhibits a much more complicated dependence of the field distribution on the frequency, and thus to describe the Stokes or anti-Stokes emission form the Raman dipole, one requires a full characterization offered by the Green's function of the system.





\section{Readout of the state}\label{sec:app_readout}

As discussed in the main text, the coupling between optical and mechanical degrees of freedom is mediated by the position operator $\hat{x}$. We use it as an observable which carries information about the excitations, and non-classical nature of the mechanical state, much like the electric field operator used to define the spectrum and statistics of emission from the system. We thus define
\begin{equation}
    {n}_x = \xzpf ^{-2} \mean{\hat{x}^{(-)} \hat{x}^{(+)}},
\end{equation}
\begin{equation}
    {G}^{(2)}_x(\tau) = \xzpf^{-4} \mean{\hat{x}^{(-)}(0) \hat{x}^{(-)}(\tau) \hat{x}^{(+)}(\tau) \hat{x}^{(+)}(0)},
\end{equation}
and the intensity correlations as
\begin{equation}
    g^{(2)}_x(\tau) = \frac{{G}^{(2)}_x(\tau)}{n_x^2}.
\end{equation}
For the mechanical system in the mixed state $\rho=\sum_i p_i \ket{\phi_i}\bra{\phi_i}$, we can express the two magnitudes as
\begin{equation}\label{eq:def.population.app}
    n_x = x_\text{zpf}^{-2} \sum_{i} p_i\sum_{k; k<i} x_{ik} x_{ki},
\end{equation}
\begin{equation}\label{eq:def.correlation.app}
    G^{(2)}_x(0) = x_\text{zpf}^{-4} \sum_{i} p_i\sum_{k; k<i} \sum_{l; l<k}\sum_{m; m>l} x_{ik} x_{kl} x_{lm} x_{mi}.
\end{equation}

In the limit of harmonic potential, the eigenstates turn into Fock states, operators $\hat{x}^{(+)}\propto \hat{b}$, $\hat{x}^{(-)}\propto \hat{b}^\dagger$, and $\hat{x}^{(0)}$ vanishes, and these magnitudes simplify to ${n}_x \rightarrow \mean{\hat{b}^\dag \hat{b}}$ and ${G}^{(2)}_x(0) \rightarrow \mean{\hat{b}^\dag \hat{b}^\dag \hat{b}\hat{b}}$. 




\section{Solving the rate equations}
\label{sec:appendix_populations}

Let us recall the definitions of rates given in Eq.~\eqref{eq:rates_thermal}, and the final form of the master equation (Eq.~\ref{eq:me2}), to write down the rate equations for the populations $p_k$ which define the mixed state density matrix $\rho =\sum_k p_k\ket{\phi_k}\bra{\phi_k}$:
\begin{align}
    \frac{\mathrm{d}}{\mathrm{d}t}p_k = &\left|\frac{x_{k,k+1}}{\xzpf}\right|^2\bar{\Gamma}_-^{(k+1)}p_{k+1} - \left|\frac{x_{k-1,k}}{\xzpf}\right|^2 \bar{\Gamma}_-^{(k)}p_k \nonumber \\ 
    &+ \left|\frac{x_{k,k-1}}{\xzpf}\right|^2 \bar{\Gamma}_+^{(k-1)}p_{k-1} - \left|\frac{x_{k+1,k}}{\xzpf}\right|^2 \bar{\Gamma}_+^{(k)}p_k.
\end{align}
For reference, for the harmonic system, where $x_{k+1,k}/\xzpf = \sqrt{k+1}$ and $x_{k,k+1}/\xzpf = \sqrt{k+1}$, and $\bar{\Gamma}_-^{(k)}$ are independent of $k$, the above simplifies to
\begin{align}
    \frac{\mathrm{d}}{\mathrm{d}t}p_k = &(k+1)\bar{\Gamma}_-p_{k+1} - \left[k \bar{\Gamma}_-+(k+1) \bar{\Gamma}_+\right]p_k \nonumber \\ 
    &+ k \bar{\Gamma}_+p_{k-1}.
\end{align}
We will now construct an algebraic formulation of the problem of finding the steady-state solution to the finite set of equations for $\left\{p_0, p_1, ..., p_K\right\}$. In the equation for $p_0$ we drop the two terms that describe emission from, and excitation to $\ket{\phi_0}$:
\begin{equation}
    \frac{\mathrm{d}}{\mathrm{d}t}p_0 = \left|\frac{x_{0,1}}{\xzpf}\right|^2\bar{\Gamma}_-^{(1)}p_{1} - \left|\frac{x_{1,0}}{\xzpf}\right|^2 \bar{\Gamma}_+^{(0)}p_0.
\end{equation}
Finally, we can turn these coupled equations into an inhomogeneous system by replacing the equation for $\dot{p}_K$ with the condition $\sum_k p_k=1$. This system of ODEs can be expressed as
\begin{equation}
    \frac{\mathrm{d}}{\mathrm{d}t} {\mathbf{v}} = M{\mathbf{v}} + \mathbf{b},
\end{equation}
with $M$ defined as
\begin{widetext}
\begin{equation}
    \xzpf^{-2}\begin{pmatrix}
    - |x_{1,0}|^2\bar{\Gamma}_+^{(0)} & |x_{0,1}|^2\bar{\Gamma}_-^{(1)} & 0 & ... & 0 & 0\\
    |x_{1,0}|^2 \bar{\Gamma}_+^{(0)} & - \left(|x_{0,1}|^2 \bar{\Gamma}_-^{(1)}+|x_{2,1}|^2\bar{\Gamma}_+^{(1)}\right) & |x_{1,2}|^2\bar{\Gamma}_-^{(2)} & ... & 0 & 0 \\
    ... \\
    0 & 0 & 0 & ... & - \left(|x_{K-2,K-1}|^2\bar{\Gamma}_-^{(K-1)}+|x_{K,K-1}|^2\bar{\Gamma}_+^{(K-1)}\right) & |x_{K-1,K}|^2\bar{\Gamma}_-^{(K)} \\
    1 & 1 & 1 & ... & 1 & 1 
    \end{pmatrix},
\end{equation}
\end{widetext}
the vector of variables $\mathbf{v} = (p_0,p_1,...,p_K)^T$ and inhomogeneous term $\mathbf{b} = (0,0,...,0,-1)^T$. Again, we can simplify it in the harmonic case to
\begin{equation}
    \begin{pmatrix}
    - \bar{\Gamma}_+ & \bar{\Gamma}_-  & ... & 0 & 0\\
    \bar{\Gamma}_+ & - [\bar{\Gamma}_-+2\bar{\Gamma}_+]  & ... & 0 & 0 \\
    ... \\
    0 & 0  & ... & - [(K-1)\bar{\Gamma}_-+K\bar{\Gamma}_+] & K\bar{\Gamma}_- \\
    1 & 1  & ... & 1 & 1 
    \end{pmatrix}.
\end{equation}

\subsection{Antibunching}\label{subsec:antibunching_app}

In the limit of 3 states only, we can find the solution as
\begin{equation}
    p_0 = \frac{\bar{\Gamma}_-^{(1)}\bar{\Gamma}_-^{(2)}}{\bar{\Gamma}_-^{(1)}\bar{\Gamma}_-^{(2)}+\bar{\Gamma}_-^{(2)}\bar{\Gamma}_+^{(0)}+\bar{\Gamma}_+^{(0)}\bar{\Gamma}_+^{(1)}},
\end{equation}
\begin{equation}
    p_{1} = \frac{\bar{\Gamma}_-^{(2)}\bar{\Gamma}_+^{(0)}}{\bar{\Gamma}_-^{(1)}\bar{\Gamma}_-^{(2)}+\bar{\Gamma}_-^{(2)}\bar{\Gamma}_+^{(0)}+\bar{\Gamma}_+^{(0)}\bar{\Gamma}_+^{(1)}},
\end{equation}
\begin{equation}
    p_{2} = \frac{\bar{\Gamma}_+^{(0)}\bar{\Gamma}_+^{(1)}}{\bar{\Gamma}_-^{(1)}\bar{\Gamma}_-^{(2)}+\bar{\Gamma}_-^{(2)}\bar{\Gamma}_+^{(0)}+\bar{\Gamma}_+^{(0)}\bar{\Gamma}_+^{(1)}},
\end{equation}
leading to the following expressions for the mechanical populations $n_x$ and intensity correlations $g_x^{(2)}(0)$:
\begin{align}
    &n_x = \xzpf^{-2}\left[p_1|x_{1,0}|^2 + \left(|x_{2,1}|^2 + |x_{2,0}|^2\right)p_0\right], \\ 
    &g_x^{(2)}(0) = \frac{p_2 |x_{2,1}x_{1,0}|^2}{n_x^2}.
\end{align}
Under harmonic approximation for matrix elements only, the expression for the populations simplifies to
\begin{equation}\label{eq:population_app}
    n_x = \frac{\bar{\Gamma}_+^{(0)}\left(\bar{\Gamma}_-^{(2)}+2\bar{\Gamma}_+^{(1)}\right)}{\bar{\Gamma}_-^{(1)}\bar{\Gamma}_-^{(2)}+\bar{\Gamma}_+^{(0)}\left(\bar{\Gamma}_-^{(2)}+\bar{\Gamma}_+^{(1)}\right)},
\end{equation}
and is plotted in Fig.~\ref{fig:spectra_nonclassical}(d) with dashed line. 

If we further assume that the anti-Stokes transition are far detuned from the optical Fano feature, and so their rates are identical ($\bar{\Gamma}_-^{(k)} \approx \bar{\Gamma}_-^{(1)}=\bar{\Gamma}_-$), we find
\begin{equation}
    n_x \approx \frac{\bar{\Gamma}_+^{(0)}\left(\bar{\Gamma}_-+2\bar{\Gamma}_+^{(1)}\right)}{\left(\bar{\Gamma}_-\right)^2+\bar{\Gamma}_+^{(0)}\left(\bar{\Gamma}_-+\bar{\Gamma}_+^{(1)}\right)}.
\end{equation}
Finally, we note that under weak pumping, the anti-Stokes emission rate should far exceeds the second Stokes transition rate ($\bar{\Gamma}_-\gg \bar{\Gamma}_+^{(1)}$),
\begin{equation}
    n_x \approx \frac{\bar{\Gamma}_+^{(0)}}{\bar{\Gamma}_-+\bar{\Gamma}_+^{(0)}} \approx \frac{{\Gamma}_+^{(0)}+\gamma \nbth}{{\Gamma}_-+\gamma+{\Gamma}_+^{(0)}}.
\end{equation}

Similarly, within the approximations of the harmonic matrix elements, and constant anti-Stokes rates, we can transform the expression for the intensity correlations
\begin{align}
    g_x^{(2)}(0)&\approx \frac{2p_k}{(p_1+2p_2)^2} \nonumber \\
    &= \frac{2\bar{\Gamma}_+^{(1)}\left[\bar{\Gamma}_-^{(1)}\bar{\Gamma}_-^{(2)}+\bar{\Gamma}_+^{(0)}\left(\bar{\Gamma}_-^{(2)}+\bar{\Gamma}_+^{(1)}\right)\right]}{\bar{\Gamma}_+^{(0)}\left(\bar{\Gamma}_-^{(2)}+2\bar{\Gamma}_+^{(1)}\right)^2}\\
    &\approx \frac{2\bar{\Gamma}_+^{(1)}\left[\left(\bar{\Gamma}_-\right)^2+\bar{\Gamma}_+^{(0)}\left(\bar{\Gamma}_-+\bar{\Gamma}_+^{(1)}\right)\right]}{\bar{\Gamma}_+^{(0)}\left(\bar{\Gamma}_-+2\bar{\Gamma}_+^{(1)}\right)^2}\\
    &\approx \frac{2\bar{\Gamma}_+^{(1)}}{\bar{\Gamma}_+^{(0)}} \left(1 + \frac{\bar{\Gamma}_+^{(0)}}{\bar{\Gamma}_-}\right) \approx \frac{2{\Gamma}_+^{(1)}}{{\Gamma}_+^{(0)}} \left(1 + \frac{{\Gamma}_+^{(0)}}{\Gamma_- + \gamma}\right).
\end{align}
In the last line, we used the definition of $\bar{\Gamma}_{\pm}^{(k)}$, and assumed negligible contribution from the thermal population $\nbth\ll 1$. Plugging in the definitions of the optomechanical rates from Eq.~\eqref{eq:def.rates}, we find the expressions given in Eq.~\eqref{eq:g2.approx}.

As the thermal population $\nbth$ increases, the last approximations break down rapidly, since the two contributions to $\bar{\Gamma}_+^{(k)}$: the thermal $\gamma\nbth$ and optomechanical ${\Gamma}_+^{(k)}$, become comparable.

\subsection{Correction to the rate equations}\label{sec:corrected.rate.equations}

Using definitions of the Stokes and anti-Stokes transition rates from Eqs.~\eqref{eq:Gamma.plus.minus.def} we can rewrite the rate equations for the mechanical population $\mean{n}=n_x$ as
\begin{align}
    \frac{\textrm{d}}{\textrm{d}t} n_x = \sum_k &k \frac{\textrm{d}}{\textrm{d}t} \rho_k \nonumber \\
    \approx \sum_k & \left\{k(k+1)\rho_{k+1} \left[\bar{\Gamma}_- - \eta_- g^2 (k+1) 2 \delta \omega_b \right] \right.\nonumber \\ 
    &- \left. k^2\rho_{k} \left[\bar{\Gamma}_- - \eta_- g^2 k 2 \delta \omega_b\right] \right.\nonumber \\
    &- \left. k(k+1)\rho_{k} \left[\bar{\Gamma}_+ + \eta_+ g^2 (k+1) 2\delta \omega_b\right] \right.\nonumber \\
    &+ \left. k^2\rho_{k-1} \left[\bar{\Gamma}_+ + \eta_+ g^2 k 2\delta \omega_b\right]
    \right\} \nonumber \\
    =-&n_x\left[\gamma+\Gamma_--\Gamma_+\right]+\Gamma_++\gamma \nbth\nonumber \\
    +&g^2 2\delta \omega_b \sum_k \rho_k \left[ k^2\eta_- + 2k^2 \eta_+ + 3 k\eta_+ + \eta_+\right]\nonumber \\
    =-&n_x\left[\gamma+\Gamma_--\Gamma_+\right]+\Gamma_++\gamma \nbth\nonumber \\
    +&g^2 2\delta \omega_b \left[ (\eta_-+2\eta_+)\mean{n^2} + 3\eta_+n_x + \eta_+\right].
\end{align}
If we approximate the state of the system as a thermal one, we can use $\mean{n^2}=n_x + 2n_x^2$, and write down the rate equation as given in Eq.~\eqref{eq:rate.equation}.





\section{Optical spectrum}\label{sec:spectra}

For the optical spectrum of a hybrid resonator, discussed in Section~\ref{sec:nonclassical}, formed by two coupled resonators characterized by frequencies $(\omega_1,\omega_2)$, dissipation rates $(\kappa_1,\kappa_2)$, and coupling $f$, we use the following expression:
\begin{equation}
    S_\text{opt}(\omega) = \Im\left[ \frac{\omega-\omega_1-i\kappa_1/2}{(\omega-\omega_1-i\kappa_1/2)(\omega-\omega_2-i\kappa_2/2) - f^2}\right].
\end{equation}
For a single mode of a plasmonic cavity, used when discussing amplification and lasing in Section~\ref{sec:amplification.lasing}, we use a simpler expression 
\begin{equation}
    S_\text{opt}(\omega) = \Im\left[ \frac{1}{\omega-\omega_1-i\kappa_1/2}\right] = \frac{\kappa_1/2}{(\omega-\omega_1)^2+(\kappa_1/2)^2}.
\end{equation}
Throughout the paper, we set the parameters to $(\omega_1, \omega_2,\kappa_1,\kappa_2, f)/2\pi = (550, 486, 60, 0.15, 15)$~THz.

We note that setup in which the laser is red-detuned from a resonant frequency of the single cavity mode is not typically used for driving the strong Stokes response (in particular, here we find that the first anti-Stokes emission rate is comparable with the first Stokes emission rate), and was chosen here to accommodate the required asymmetry of the Fano feature. Nevertheless, the mechanical system becomes excited through the Stokes transitions, in a quantum-mechanical analogue of the vibrational pumping mechanism.

\section{Dependence on the laser frequency}\label{app:laser.freq}

In Fig.~\ref{fig:spectra_nonclassical}(f,g) we showed how sub-Poissonian statistics of the mechanical state depends on the anharmonicity $\delta \omega_b$ and the coherent cavity population $|\alpha|^2$ for a particular laser frequency of 501~THz. In particular, we showed that the system exhibits maximally nonclassical response for $\omega_l-(\tilde{\omega}_2-\tilde{\omega}_1)$ tuned to the dip of the Fano feature in the optical spectrum $S_\text{opt}$. 

If we wanted to further decrease $g_x^{(2)}(0)$, we could explore larger effective optomechanical interaction to dominate over the thermal response, by increasing $|\alpha|^2$. However, we can clearly see from  Fig.~\ref{fig:spectra_nonclassical}(g) that $g_x^{(2)}(0)$ increases quickly with $|\alpha|^2>2$. This is because the Stokes transitions $\omega_l-(\tilde{\omega}_2-\tilde{\omega}_1)$ become detuned from the Fano dip.

\begin{figure}
    \includegraphics[width=\linewidth]{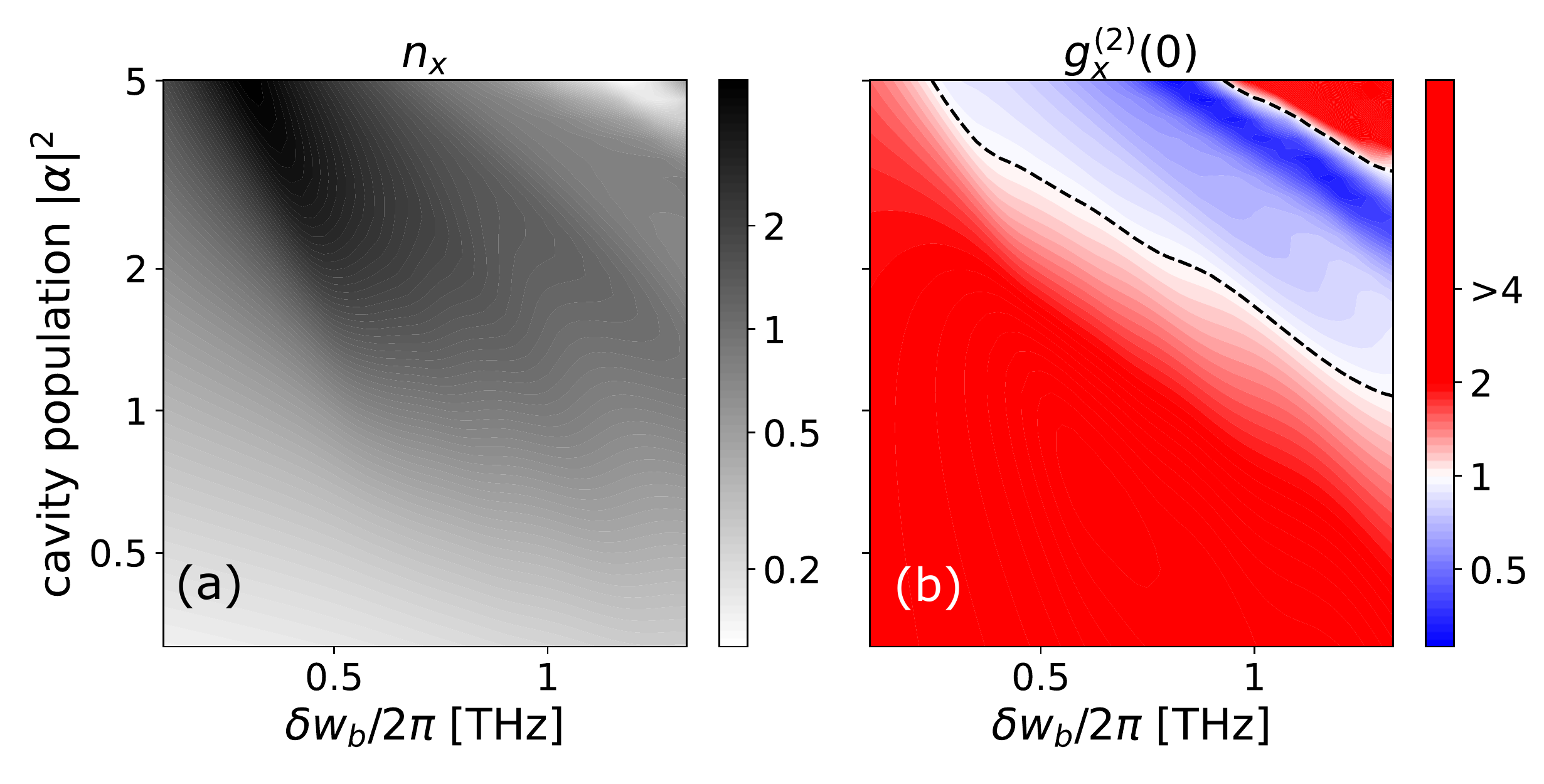}
    \caption{(a) Phonon populations (see Eq.~\eqref{eq:phonon_population}) and (b) intensity correlations (Eq.~\eqref{eq:correlations}) for the laser frequency set to 495~THz (compared to 501~THz used in Fig.~\ref{fig:spectra_nonclassical}), as a function of the anharmonicity $\delta \omega_b$ and cavity population $|\alpha|^2$. Remaining parameters are chosen as Fig.~\ref{fig:spectra_nonclassical}.}
    \label{fig:spectra_nonclassical_app}
\end{figure}

To achieve a stronger degree of sub-Poissonian statistics, in Fig.~\ref{fig:spectra_nonclassical_app} we consider a setup with the slightly lower laser frequency (496~THz), for which $\omega_l-(\tilde{\omega}_2-\tilde{\omega}_1)$ should be tuned to the Fano dip at larger values of $|\alpha|^2$. Indeed, we find that $g_x^{(2)}(0)$ saturates at the significantly lower values of $0.2$ at larger cavity populations.

\section{Dependence on the thermal equilibrium population}\label{app:temp}

\begin{figure}
    \includegraphics[width=\linewidth]{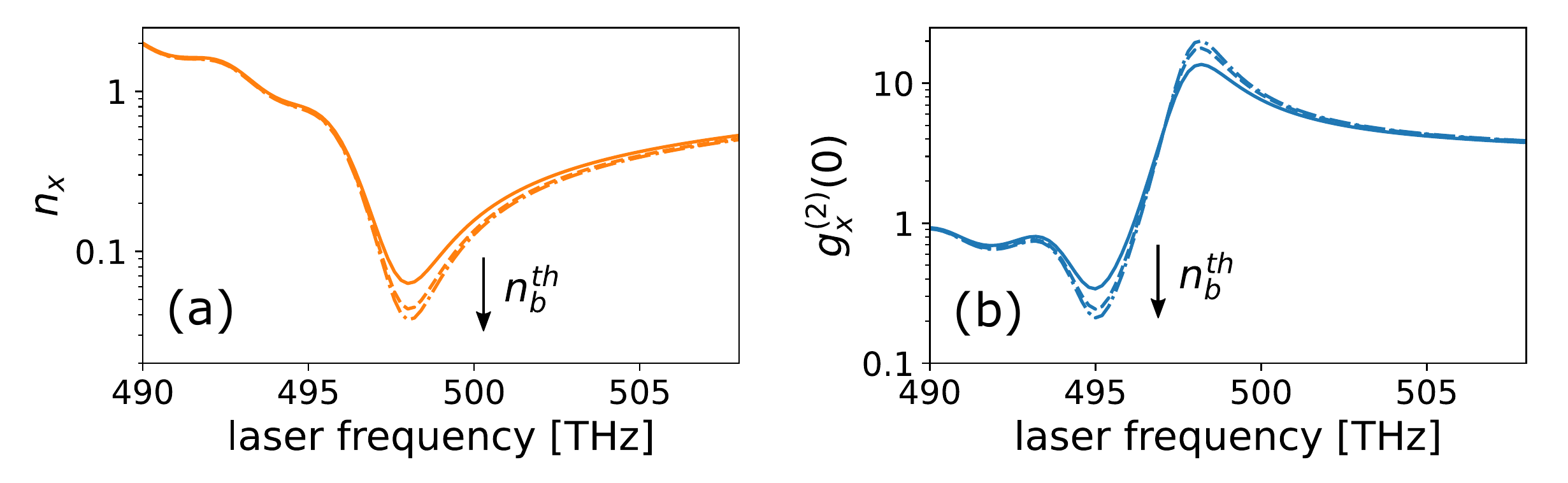}
    \caption{(a) Mechanical populations $n_x$ (see Eq.~\eqref{eq:phonon_population}) and (b) intensity correlations (Eq.~\eqref{eq:correlations}) for decreasing population of thermal phonons in the bath: $\nbth=0.05$ (solid lines), $\nbth=0.02$ (dashed lines), $\nbth=0.01$ (dashed-dotted lines). Remaining parameters are chosen as Fig.~\ref{fig:spectra_nonclassical}.}
    \label{fig:thermal_app}
\end{figure}

In Fig.~\ref{fig:thermal_app} we show how the minima of the mechanical populations and intensity correlations become more pronounced as we decrease the thermal populations $\nbth$. 

\section{Implementation of the classical trajectories}\label{app:trajectories}

Equations~\eqref{eq:trajectories.harmonic} can be rewritten into a form more suitable for numerical solution by normalizing coordinates $\tilde{x} = x/\xzpf$, $\tilde{p} = p \xzpf/\hbar$, and introducing time parameter $\tau=t \omega_b$:
\begin{subequations}    
    \begin{equation}
        \frac{\mathrm{d}}{\mathrm{d}\tau} \alpha = -i\frac{\Delta}{\omega_b} \alpha - i\frac{g_0}{\omega_b} \tilde{x}\alpha - i\frac{\Omega}{\omega_b} - \frac{\kappa}{2\omega_b} \alpha,\label{eq:tildea}
    \end{equation}
    \begin{equation}
        \frac{\mathrm{d}}{\mathrm{d}\tau} \tilde{x} = 2 \tilde{p} -  \frac{\gamma}{2\omega_b} \tilde{x},\label{eq:tildex}    
    \end{equation}
    \begin{equation}
        \frac{\mathrm{d}}{\mathrm{d}\tau} \tilde{p} = -\frac{g_0}{\omega_b} |\alpha|^2 - \frac{1}{2}\tilde{x}-  \frac{\gamma}{2\omega_b} \tilde{p}.\label{eq:tildep}    
    \end{equation}
\end{subequations}

For an Morse potential, we modify Eq.~\eqref{eq:tildep} to read
\begin{equation}\label{eq:tildep.2}
    \frac{\mathrm{d}}{\mathrm{d}\tau} \tilde{p} = -\frac{g_0}{\omega_b} |\alpha|^2 - \frac{1}{2\tilde{a}}\left(1-e^{-\tilde{a}\tilde{x}}\right)e^{-\tilde{a}\tilde{x}}-  \frac{\Gamma}{2\omega_b} \tilde{p},
\end{equation}
where we introduced $\tilde{a}=a \xzpf$. Notably, as $\tilde{a}$ approaches 0, we recover the harmonic restoring force.

\subsection{Mechanical lasing amplitudes}\label{app:lasing}

\begin{figure}
    \includegraphics[width=\linewidth]{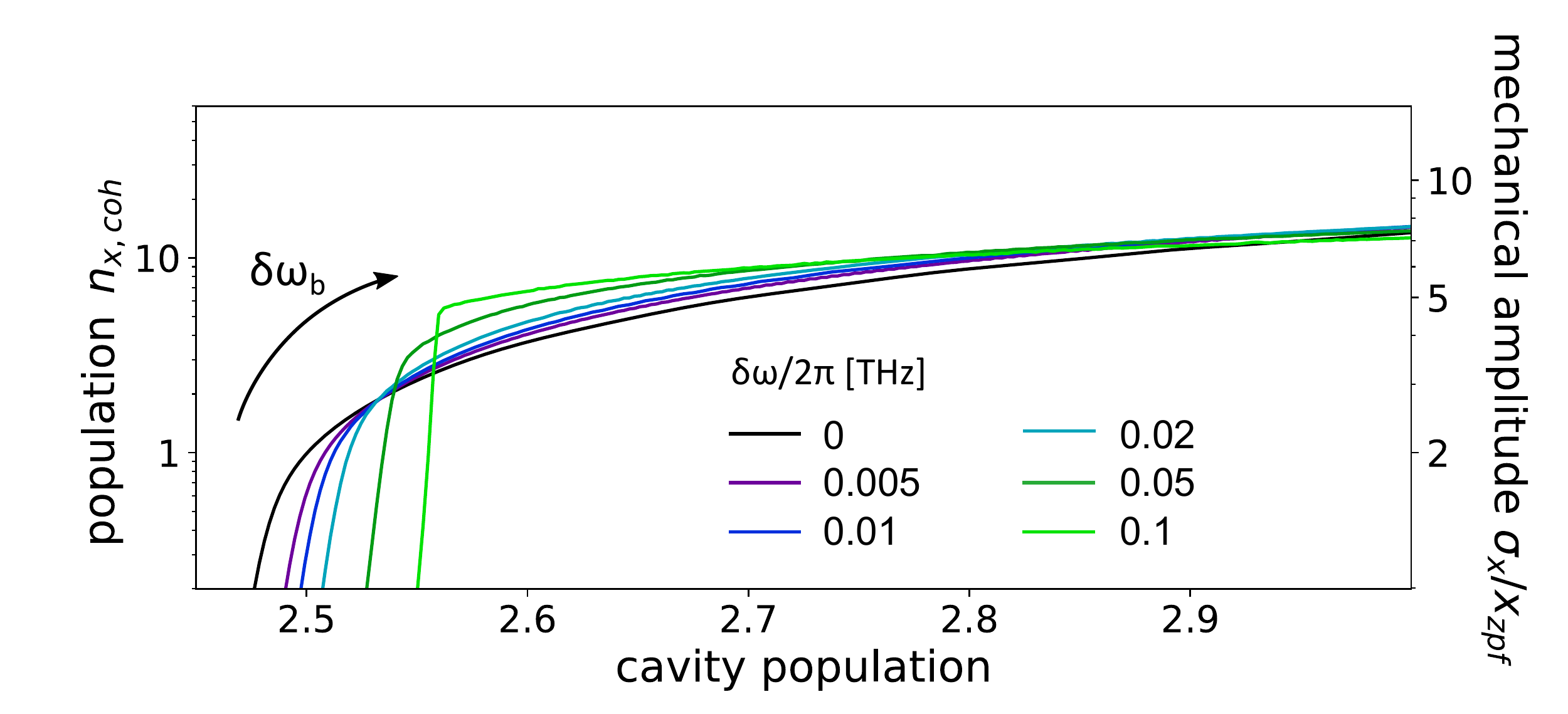}
    \caption{Dependence of the amplitude of the coherent mechanical oscillations on the anharmonicity $\delta \omega_b$, and populations of the optical cavity near the threshold. As in Fig.~\ref{fig:results_lasing}(c), we quantify the oscillations through the equivalent coherent population $n_{x,\text{coh}}$, and the normalized standard deviation $\sigma_x/\xzpf$.}
    \label{fig:results_lasing_app}
\end{figure}

In Fig.~\ref{fig:results_lasing_app} we revisit the results included in Fig.~\ref{fig:results_lasing}(c), to show how the amplitudes of mechanical lasing change with small anharmonicity. We note that the threshold clearly shifts towards larger cavity populations (included in the definition of the effective optomechanical coupling $g$), and the oscillations becomes initially more steep.

\bibliography{references_hardcopy}

\end{document}